\documentclass{INTERSPEECH2023}
\usepackage{hyperref}
\usepackage[usenames,dvipsnames]{xcolor}  % colors
\usepackage{multirow}
\usepackage{balance}
\interspeechcameraready

\title{Target Active Speaker Detection with Audio-visual Cues}
\name{Yidi Jiang$^1$, Ruijie Tao$^1$, Zexu Pan$^1$, Haizhou Li$^{1,2}$}
\address{
  $^1$National University of Singapore, Singapore\\
  $^2$Shenzhen Research Institute of Big Data, School of Data Science, \\
  The Chinese University of Hong Kong, Shenzhen, China
  }
\email{\{yidi\_jiang,ruijie.tao,pan\_zexu\}@u.nus.edu, haizhouli@cuhk.edu.cn}

\begin{document}
\maketitle
 
\begin{abstract}
In active speaker detection (ASD), we would like to detect whether an on-screen person is speaking based on audio-visual cues. Previous studies have primarily focused on modeling audio-visual synchronization cue, which depends on the video quality of the lip region of a speaker. In real-world applications, it is possible that we can also have the reference speech of the on-screen speaker. To benefit from both facial cue and reference speech, we propose the Target Speaker TalkNet (TS-TalkNet), which leverages a pre-enrolled speaker embedding to complement the audio-visual synchronization cue in detecting whether the target speaker is speaking.
Our framework outperforms the popular model, TalkNet on two datasets, achieving absolute improvements of 1.6\% in mAP on the AVA-ActiveSpeaker validation set, and 0.8\%, 0.4\%, and 0.8\% in terms of AP, AUC and EER on the ASW test set, respectively. Code is available at \href{https://github.com/Jiang-Yidi/TS-TalkNet/}{\textcolor{Blue}{https://github.com/Jiang-Yidi/TS-TalkNet/}}.
%To the best of our knowledge, this is the first work that demonstrates the benefits of incorporating speaker characteristics for ASD. 

\end{abstract}

\vspace{-0.2cm}
\section{Introduction}
The study of active speaker detection (ASD) is to determine whether an on-screen person is speaking in each video frame of an audio-visual scene~\cite{roth2020ava}. It plays a critical frontend role for various speech processing tasks, such as audio-visual speaker localization~\cite{qian2021multi}, speaker verification~\cite{tao2022self,liu2022mfa,liu2022neural}, speaker extraction~\cite{ochiai2019multimodal,pan2022seg}, speech recognition~\cite{afouras2018deep,wang2022predict},
among others~\cite{yang2022KF,yang2022deep}.
%, as these algorithms are usually applied to speech signals only.

% ASD methods usually learn audio-visual synchronization, but this highly rely on the accurate lip region.
Typically, ASD methods rely on modeling the synchronization between the audio and visual modalities to detect whether a person is talking~\cite{pouthier2021active,alcazar2021maas,zhang2019multi}. Early approaches focused on short temporal segments of speech activity~\cite{ding2020personal,medennikov2020target}, lip movement~\cite{patrona2016visual} and audio-visual synchronization~\cite{afouras2020self, chung2019naver, chung2017out} to identify the active speaker. These methods have an efficient structure, but lack the ability to capture the long-term temporal association between the audio and visual signals. The recent work, TalkNet~\cite{tao2021someone}, incorporated long-term temporal information and demonstrated strong performance on the ASD task. Following this pipeline, other works attempted to model audio-visual synchronization in a more efficient way, such as Graph Neural Networks~ (GNN)~\cite{alcazar2022end, min2022learning}. 
However, obtaining high-resolution video of the lip regions is not always feasible, which can hinder the accurate determination of audio-visual synchronization.
%However, high-resolution video of the lip region is needed to determine accurate audio-visual synchronization, which is not always possible.
% in complex acoustic environments, accurate synchronization information can be difficult to detect, which still limits the application scenarios of ASD systems.

% Motivation: How human did, how others did
Human has an inherent ability to focus on a known speaker in complex social environments, known as selective auditory attention in a ``cocktail party''~\cite{bronkhorst2000cocktail,bronkhorst2015cocktail,mesgarani2012selective}. In other words, when we are already familiar with a person's voice, i.e., the target speaker, we can perform selective listening by comparing the heard voice with the reference voice.

In target speaker extraction task, the voice of a specific speaker is pre-enrolled in form of a speaker embedding. This speaker embedding is then used as a reference for target speaker extraction~\cite{ge2020spex+, pan2021reentry}. 
Similarly, target speaker voice activity detection maps unknown speech to extracted target speaker embeddings to perform speaker diarization task~\cite{ding2020personal,chen2023attention}. In~\cite{chung2019said}, speaker diarization task was reformulated to a supervised classification problem by leveraging audio-visual correlation and speaker models for each speaker.
These studies highlight the benefits of understanding a target speaker's voice characteristic in speech processing.
Motivated by these studies, we believe that the reference speech could be informative in ASD task as well, in particular, in challenging acoustic environments. %However, this approach has been ignored by current studies.
% In addition to audio-visual synchronization, speaker embedding is an essential cue for solving ASD problems under challenging conditions. Given the success of neural network-based speaker recognition models~\cite{desplanques2020ecapa,bai2021speaker}, incorporating speaker embeddings can potentially improve the performance of ASD systems.

In this work, we propose the Target Speaker TalkNet (TS-TalkNet) framework as an extension of TalkNet~\cite{tao2021someone}, introducing the concept of the ``target speaker" to the ASD task. TS-TalkNet first enrolls the on-screen speaker as the target speaker and then performs auditory attention with the corresponding pre-enrolled speaker embedding for the ASD task. To achieve this, we construct a face-speaker library where each target speaker's face is associated with their pre-enrolled reference speech (if exists), providing the target speaker embedding. TS-TalkNet explores the voice characteristic cue to determine whether the audio signals resemble the target speaker embedding, which complements the audio-visual synchronization cue. Noted that if reference speech does not exist, TS-TalkNet only utilizes the synchronization cue. We also explore various multi-modal fusion methods to investigate the efficacy of leveraging speaker embedding in TS-TalkNet. The contributions of this paper can be summarised as follows:

\begin{enumerate}
    \item We explore the use of reference speech to assist active speaker detection.
    \item We propose the TS-TalkNet framework that fully makes use of both audio and visual cues, i.e., voice characteristic signal and audio-visual synchronization information. 
    
    %, and verify that voice characteristic is an important cue for ASD study.
    \item We conduct experiments on the AVA-ActiveSpeaker~(AVA) and the Active Speakers in the Wild~(ASW) datasets with improvements: 1.6\% in mAP on the AVA validation set; 0.8\%, 0.4\%, and 0.8\% in AP, AUC and EER, respectively, on the ASW test set, confirming the superiority of TS-TalkNet. 
\end{enumerate}

\begin{figure*}[!t]
% \vspace{-0.2cm}
    \centering
      \includegraphics[scale=0.85]{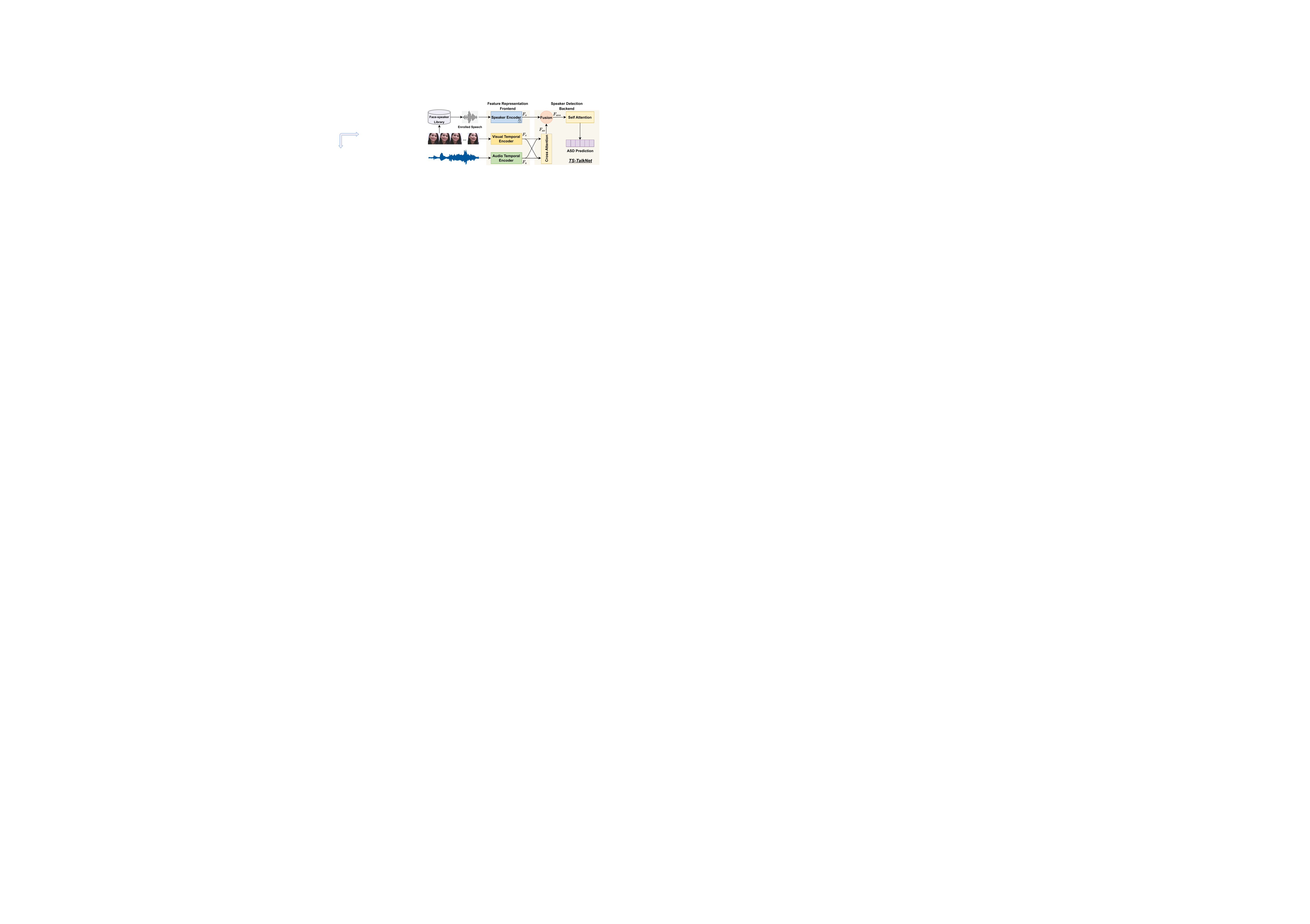}
    \vspace{-0.2cm}
    \caption{The overview framework of our TS-TalkNet. It consists of a feature representation frontend and a speaker detection backend classifier. The feature representation frontend includes audio and visual temporal encoders, and speaker encoder. The speaker detection backend comprises a cross-attention and a fusion module to combine the audio, visual and speaker embeddings, and a self-attention module to predict the ASD scores. The lock represents the speaker encoder is frozen in our framework.}
    \label{fig:overview}
    \vspace{-0.4cm}
\end{figure*}
\vspace{-0.2cm}
\section{TS-TalkNet: Target Active Speaker Detection}
%In this section, we present our proposed framework named Target Speaker TalkNet (TS-TalkNet). 
As depicted in Figure \ref{fig:overview}, our proposed TS-TalkNet framework takes the cropped face track video, corresponding audio signal and the target speaker's enrolled speech signal as inputs, and outputs the binary decision of the speech activity for each video frame. TS-TalkNet consists of a feature representation frontend and a speaker detection backend.

\vspace{-0.1cm}
\subsection{Feature representation frontend}
The feature representation frontend includes an audio temporal encoder, a visual temporal encoder and a speaker encoder. Following TalkNet, the audio and visual temporal encoders are used to learn audio and visual embeddings from the audio and face track inputs. The speaker encoder is proposed to supplement the target speaker information.
% In our proposed TS-TalkNet framework, we leverage the visual and audio temporal encoders from TalkNet.  
~\\
\textbf{Visual temporal encoder.}
The visual temporal encoder is designed to learn the long-term representations of facial expression dynamics by encoding the visual stream into a sequence of visual embeddings $F_v$ with consistent time resolution. To achieve this, the visual frontend captures spatial information within each video frame and encodes the video frame stream into a sequence of frame-based embeddings. The visual temporal network then employs a video temporal convolutional block, comprising of five residual connected ReLU, batch normalization, and depth-wise separable convolutional layers, followed by a Conv1D layer that reduces the embedding dimension. The purpose of this encoder is to capture the temporal content within a long-term visual spatio-temporal structure.
~\\
\textbf{Audio temporal encoder.}
The audio temporal encoder is a ResNet34 network~\cite{he2016deep} with a squeeze-and-excitation (SE) module~\cite{hu2018senet}. This network extracts audio content representation from the temporal dynamics of a sequence of audio frames, which are initially represented by a vector of Mel-frequency cepstral coefficients (MFCCs). The output of this network is a sequence of audio embeddings, denoted as $F_a$. $F_a$ matches the time resolution of the visual embeddings, $F_v$.
~\\
\textbf{Speaker encoder.}
To begin, we construct the face-speaker library, which contains multiple reference speeches for each target speaker. Further details regarding the construction process can be found in Subsection~\ref{subsection:library}.

We designate the individual of each face track input as the `target speaker'. Then for each face track, we randomly select one pre-enrolled speech from all the associated speeches in the face-speaker library. The pre-enrolled speech is then used as the input of the speaker encoder. To incorporate target speaker characteristics into our TS-TalkNet, we leverage a pre-trained speaker recognition model to obtain the target speaker embedding. To guarantee robust performance, the ECAPA-TDNN model~\cite{desplanques2020ecapa} is used as the speaker encoder. It has demonstrated reliable performance for speaker recognition task.

The ECAPA-TDNN model employs emphasized channel attention to selectively focus on critical parts of the speech signal, propagating that information through the network and aggregating it to make a final decision. From the variable lengths of input utterances, the output speaker embedding has the fixed dimension. This capability enables the model to handle speech signals in various scenarios and generate robust speaker representations more effectively.

From the enrolled speech input, we obtain the target speaker embedding $F_s$ and convey the target speaker information to the speaker detection backend. We freeze the parameters of this module in our framework, since our purpose is to obtain robust embedding for the target speaker.
\vspace{-1mm}
\subsection{Speaker detection backend}
\label{subsection:fusion}
The speaker detection backend comprises a cross-attention module to achieve audio-visual synchronization, a fusion module to combine the speaker embedding with audio-visual embeddings, in addition to a self-attention module to capture speaking activities from the temporal context at the utterance level.
We aims to combine three embeddings generated from the feature representation frontend: $F_a$ for audio modality, $F_v$ for visual modality, and $F_s$ for speaker characteristic, to provide a comprehensive representation of the speaking activities for ASD prediction.
~\\
\textbf{Cross-attention module.} Firstly, we use a cross-attention structure along the temporal dimension to achieve audio-visual synchronization and interaction.
The inputs are the vectors of query ($Q_a$, $Q_v$), key ($K_a$, $K_v$), and value ($V_a$, $V_v$) from audio and visual embeddings, $F_a$ and $F_v$ respectively, projected by a linear layer. As formulated in Equation~\ref{eq:a-v} and Equation~\ref{eq:v-a}, the outputs are the audio attention embedding $F_{a \rightarrow v}$ and visual attention embedding $F_{v \rightarrow a}$, where $d$ denotes the dimension of $Q$, $K$and $V$.
The outputs are concatenated together along the temporal direction as audio-visual embedding $F_{av}$.
\vspace{-2mm}
\begin{equation}
    F_{a \rightarrow v} = softmax(\frac{Q_vK_a^\mathrm{T}}{\sqrt{d}})V_a
    \label{eq:a-v}
\end{equation}
\vspace{-2mm}
\begin{equation}
    F_{v \rightarrow a} = softmax(\frac{Q_aK_v^\mathrm{T}}{\sqrt{d}})V_v
    \label{eq:v-a}
\end{equation}

\begin{figure}[!t]
    \centering
      \includegraphics[scale=0.5]{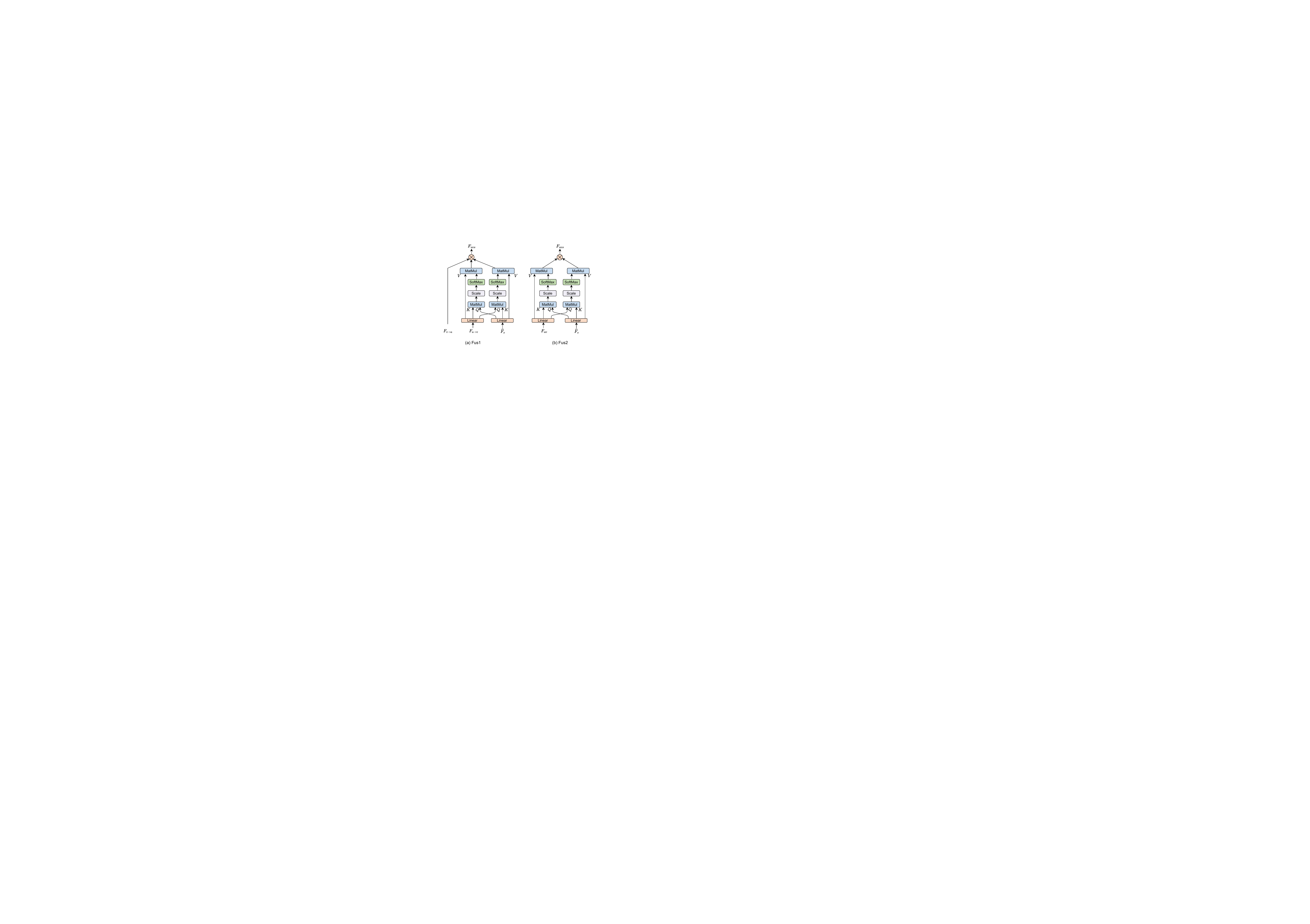}
    \vspace{-0.2cm}
    \caption{The different fusion approaches for audio, visual and speaker embeddings. 
    We denote the structure (a) and (b) as `Fus1' and `Fus2', respectively. $\otimes$ represents embedding concatenation along the time dimension.}
    \vspace{-0.4cm}
    \label{fig:fusion}
\end{figure}

\vspace{-2mm}
~\\
\textbf{Fusion module.} 
The fusion module is used to combine the speaker embedding with the audio-visual embedding to obtain the overall audio-visual-speaker representation $F_{avs}$. 

% Noted that $F_s \in \mathbb{R}^{1 \times D_s}$ is not a temporal embedding, where $D_s$ is the dimension of speaker embedding. While $F_{av} \in \mathbb{R}^{T \times D_a}$ and $F_{a \rightarrow v} \in \mathbb{R}^{T \times D_a}$ are temporal embedding sequences, where $D_a$ is the dimension of audio embedding. We just replicate $F_s$ along the time dimension to $\hat {F}_s \in \mathbb{R}^{T \times D_s}$ to implement the fusion module.

Noted that $F_s$ is not a temporal embedding, while $F_{av}$, $F_{v \rightarrow a}$ and $F_{a \rightarrow v}$ are all temporal embedding sequences. For alignment, we replicate $F_s$ along the time dimension to $\hat {F}_s$ to implement the fusion module.

We have investigated three different fusion structures to demonstrate that our idea of leveraging voice characteristics isn't restricted to a specific network architecture.
As shown in Fig~\ref{fig:fusion} (a) and (b), `Fus1' structure uses $F_{a \rightarrow v}$ and $\hat {F}_s$ as inputs for cross-attention. `Fus2' structure applies $F_{av}$ to achieve the cross-attention process with $\hat {F}_s$.
The third structure is to concatenate the $\hat {F}_s$ and $F_{av}$ along the temporal dimension, which is denoted as `Concat'.
~\\
\textbf{Self-attention module.}
After obtaining the audio-visual-speaker representation $F_{avs}$, we follow the same strategy as TalkNet~\cite{tao2021someone} and apply a self-attention structure to model $F_{avs}$ temporal information to distinguish between speaking and non-speaking frames. This structure is similar with the cross-attention structure, except that the query, key and value in the attention layer all come from the joint embedding $F_{avs}$.

\vspace{-2mm}
\subsection{Loss function}
Treating ASD as a frame-level classification task, we project the output of the self-attention structure to an ASD label sequence using a fully connected layer and the softmax operation. We then compare the predicted label sequence to the ground truth label sequence using cross-entropy loss, which is given in Equation~\ref{eq:loss}. Here, $\hat{y}_i$ and $y_i$ represent the predicted and ground truth ASD labels for the $i^{th}$ video frame, respectively, where $i \in [1, T]$. The value of $T$ represents the number of video frames.
\vspace{-2mm}
\begin{equation}
    \mathcal{L} = -\frac{1}{T}  \sum_{i=1}^{T} (y_i \cdot log \hat{y}_i) + (1-y_i) \cdot log (1- \hat{y}_i)
    \label{eq:loss}
\end{equation}
\vspace{-5mm}
\section{Experiments}
In this section, we describe the datasets and experimental details to evaluate the proposed TS-TalkNet framework.
\vspace{-2mm}
\subsection{Dataset}
\textbf{AVA-ActiveSpeaker (AVA) dataset}~\cite{roth2020ava} is a large-scale audio-visual active speaker detection dataset. 
It consists of 262 videos extracted from Hollywood movies, with 120 for training, 33 for validation, and 109 for testing.
This dataset contains about 3.65 million human-labeled video frames or 38.5 hours of face tracks, and the corresponding audio.
~\\
\textbf{The Active Speakers in the Wild (ASW) dataset}~\cite{kim2021look} consists of 212 videos randomly selected from the VoxConverse dataset~\cite{chung2020spot}. It contains 13.4 hours videos in which 56.7\% are active in the training set, 9.6 hours videos where 60.4\% are active in the validation set, and 7.9 hours videos where 57.0\% are active in the test set. 
\vspace{-1mm}
\subsection{Face-speaker library}
\label{subsection:library}
We pre-process the dataset to build the face-speaker library by searching the face tracks from the same person using a face recognition module. 
We utilize the pre-trained ResNet50 model on the Glint360K dataset~\cite{an2022pfc} as the face recognition module and set the face similarity threshold at 0.7.
The enrolled speeches come from the corresponding audio of the found tracks, which are located by extracting the active segments using ground-truth labels.
Then for each input, we can retrieve the pre-enrolled speeches of the target speaker from the face-speaker library according to the input faces.
If the reference speeches exist, we randomly choose one as the target speaker enrolled speech and feed it into TS-TalkNet. Otherwise, we set the enrolled speech to zero vectors to process detection. 
In this way, our TS-TalkNet can handle different situations, whether there is an enrolled speech record or not.
\vspace{-2mm}
\subsection{Implementation details}
The proposed TS-TalkNet is implemented by PyTorch using the Adam optimizer. We set the initial learning rate to $10^{-4}$ and decrease it by 5\% for each epoch. The MFCC dimension is set to 13, and all faces are resized into $112 \times 112$ pixels. Both audio and visual embeddings are set to 128 dimensions. For both cross-attention and self-attention structures, a single transformer layer with 8 attention heads is used. We apply visual augmentation such as random flipping, rotating, and cropping the original images to improve framework performance.
\vspace{-2mm}
\subsection{Evaluation metric}
For the AVA dataset, we use the official ActivityNet evaluation tool to compute mean average precision~(mAP) and evaluate on the validation~(val) set. For the ASW dataset, following~\cite{kim2021look,tao2021someone}, we compute three metrics: mean precision~(AP), area under the receiver operating characteristic~(AUC), and equal error rate~(EER) using sklearn package and evaluate on the ASW val and test set. 

\begin{table}[t]
\caption{Comparison with the state of the arts on the AVA val set in terms of mAP. The bold rows represent the results of our proposed methods.}
\footnotesize
\centering
  \begin{tabular}{p{4cm}<{\centering}p{2cm}<{\centering}}
  \toprule
   Method & mAP~(\%)$\uparrow$ \\
   \midrule
   Roth et al.~\cite{alcazar2021maas,roth2020ava} & 79.2 \\
   Zhang et al.~\cite{zhang2019multi} & 84.0 \\
   MAAS-LAN~\cite{alcazar2021maas} & 85.1 \\
   Alcazar et al.~\cite{alcazar2020active} & 87.1 \\
   Chung et al~\cite{chung2019naver} & 87.8 \\
   MAAS-TAN~\cite{alcazar2021maas} & 88.8 \\
   UniCon~\cite{zhang2021unicon} & 92.0 \\
   TalkNet~\cite{tao2021someone} &92.3 \\
   ASDNet~\cite{kopuklu2021design} &93.5 \\
   Light-ASD~\cite{liao2023light} &94.1 \\
   UniCon+~\cite{zhang2022unicon+} &94.5 \\
   SPELL+~\cite{min2022learning} &94.9\\
   \midrule
   \textbf{TS-TalkNet~(Fus1)} &\textbf{93.3}\\
   \textbf{TS-TalkNet~(Fus2)} &\textbf{93.5}\\
   \textbf{TS-TalkNet~(Concat)} &\textbf{93.9} \\
\bottomrule
     \end{tabular}
     % }
 \label{tab:main1}
 \vspace{-0.4cm}
\end{table}

\begin{table}[t!]
\caption{Comparison with the state of the arts on the ASW val and test set in terms of AP, AUC and EER. The bold rows represent the results of our proposed methods.}
\scriptsize
\centering
% {
  % \setlength{\tabcolsep}{1.5mm}{
  % \begin{tabular}{c|c|ccc}
  \begin{tabular}{p{.3cm}<{\centering}|p{2.4cm}<{\centering}|p{1cm}<{\centering}p{1cm}<{\centering}p{1cm}<{\centering}}
  \toprule   
   Set&Method &AP~(\%)$\uparrow$ &AUC~(\%)$\uparrow$ &EER~(\%)$\downarrow$\\
   \midrule
   \multirow{4}{*}{Val}&TalkNet~\cite{tao2021someone}&96.4&98.2&6.0 \\
   &\textbf{TS-TalkNet~(Fus1)}&\textbf{97.3}&\textbf{98.6}&\textbf{5.4} \\
   &\textbf{TS-TalkNet~(Fus2)} &\textbf{97.5}&\textbf{98.7}&\textbf{5.1}\\
   &\textbf{TS-TalkNet~(Concat)}&\textbf{97.7}&\textbf{98.7}&\textbf{5.1}\\
   \midrule
   \multirow{9}{*}{Test}&RothNet~\cite{roth2020ava} &89.7&-&-\\
   &SyncNet~\cite{chung2017out} &92.4&-&-\\
   &ASW-SSL~\cite{kim2021look}  &90.5&-&-\\
   &LoCoNet~\cite{wang2023loconet} &93.4&95.1&9.8\\
   &ASW-BGRUs~\cite{kim2021look} &96.6&97.2&6.2\\
   &TalkNet~\cite{tao2021someone} &97.7&98.6&5.1 \\
   &\textbf{TS-TalkNet~(Fus1)} &\textbf{98.3}&\textbf{98.9}&\textbf{4.8}\\
   &\textbf{TS-TalkNet~(Fus2)} &\textbf{98.3}&\textbf{98.9}&\textbf{4.4}\\
   &\textbf{TS-TalkNet~(Concat)} &\textbf{98.5}&\textbf{99.0}&\textbf{4.3}\\
\bottomrule
     \end{tabular}
     % }
 \label{tab:main2}
 \vspace{-2mm}
\end{table}

\vspace{-1mm}
\section{Results and Analysis}
In this section, we present the performance comparison of our proposed TS-TalkNet with state-of-the-art methods on the AVA and ASW datasets to show its efficiency. We also explore the architectures of the fusion module and analyze the results to find out the improvement under the different test conditions. 
\vspace{-1mm}
\subsection{AVA results}
We firstly report the performance of TS-TalkNet on the AVA val dataset in Table~\ref{tab:main1}.  With the inclusion of enrolled speech, we observe that TS-TalkNet~(Concat) achieves 93.9\% mAP and 1.6\% mAP improvements over TalkNet on the AVA val set.
Moreover, for TS-TalkNet~(Fus1) and TS-TalkNet~(Fus2) frameworks, we obtain the improvements by 1.0\% and 1.2\% mAP, respectively.
\vspace{-2mm}
\subsection{ASW results}
The results of TS-TalkNet on the ASW val and test dataset are reported in Table~\ref{tab:main2}.
Specially, the proposed TS-TalkNet~(Concat) framework outperforms TalkNet by 1.3\%, 0.6\% and 0.9\% in terms of AP, AUC and EER, respectively, on the ASW val set; and 0.8\%, 0.4\% and 0.8\% in terms of AP, AUC and EER, respectively, on the ASW test set. Similarly, both TS-TalkNet~(Fus1) and TS-TalkNet~(Fus2) frameworks outperform TalkNet by a large margin.

\begin{figure}[h]
\vspace{-0.2mm}
    \centering
      \includegraphics[scale=0.36]{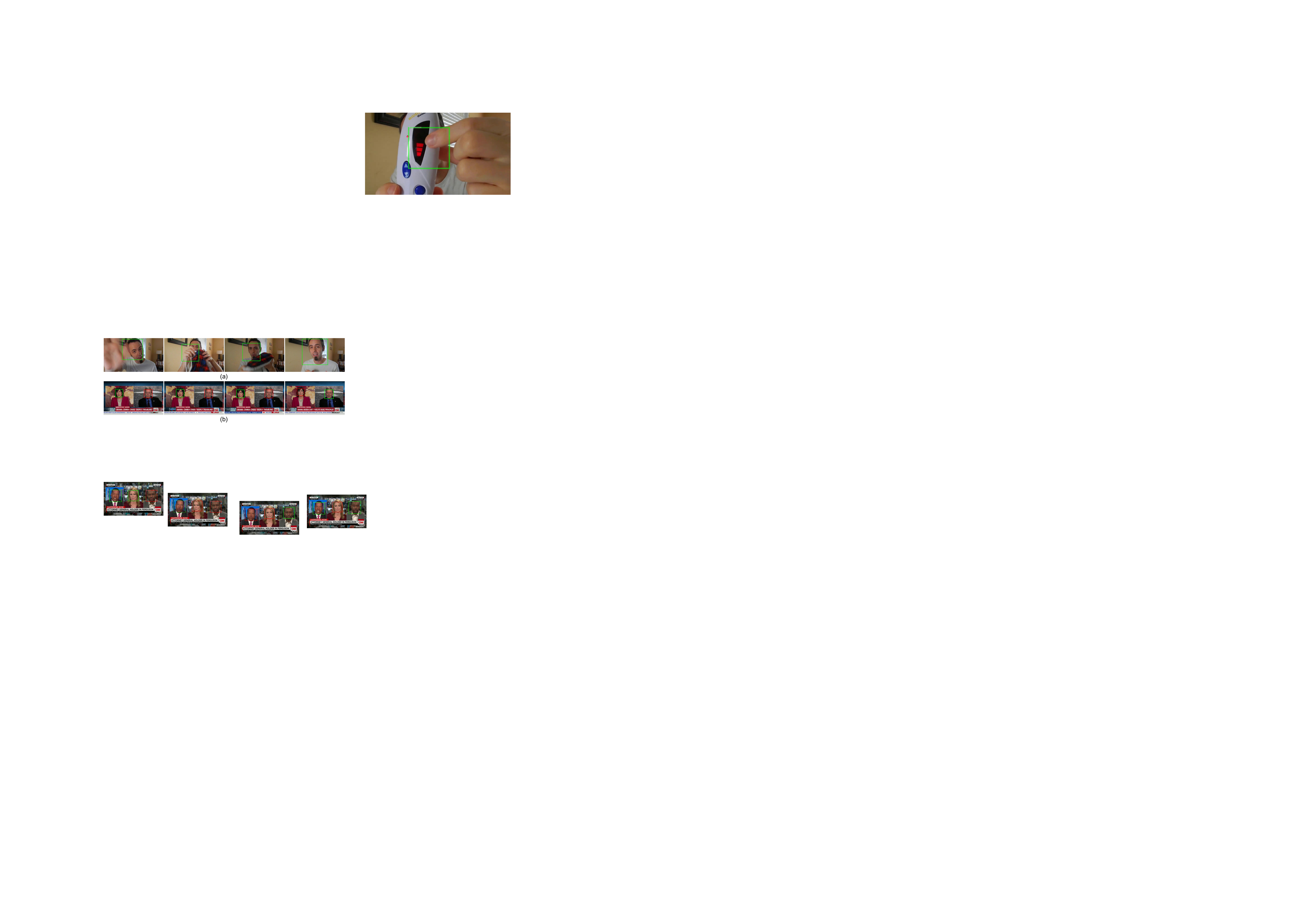}
    % \vspace{-0.1cm}
    \caption{The results of TS-TalkNet for real-world videos with one person~(a) and multiple persons~(b) on the screen. The green box denotes the active speaker. The red box denotes the inactive speaker. As demonstrated in (a), despite the occlusion of the lip in certain video frames, which compromises the audio-visual synchronization cue, TS-TalkNet can still accurately detect the speaker with the aid of speaker embedding complementation.}
    \label{fig:viz}
\end{figure}

We visualize the ASD results in Figure~\ref{fig:viz} for two videos from the ASW dataset. Our results verify the benefits of speaker embedding on ASD task and demonstrate the superiority of TS-TalkNet.

\begin{table}[t]
\caption{Results comparison of different percentages of the frames with active labels on the ASW val set. For example, `0-20' represents the tracks with less than 20\% active labels.}
\footnotesize
\centering
\setlength{\tabcolsep}{1.5mm}{
  \begin{tabular}{ccccc}
  
   \toprule
Active Frame~(\%)&Method&AP~(\%)$\uparrow$&AUC~(\%)$\uparrow$&EER~(\%)$\downarrow$\\
\midrule
\multirow{2}{*}{`0 - 20'}&TalkNet&8.29&95.92&9.45\\
&\textbf{TS-TalkNet}&\textbf{11.54}&\textbf{96.65}&\textbf{7.63}\\
\midrule
\multirow{2}{*}{`20 - 40'}&TalkNet&83.47&94.85&11.22\\
&\textbf{TS-TalkNet}&\textbf{89.37}&\textbf{96.06}&\textbf{9.41}\\
\midrule
\multirow{2}{*}{`40 - 60'}&TalkNet&92.54&93.33&13.95\\
&\textbf{TS-TalkNet}&\textbf{93.39}&\textbf{93.94}&\textbf{13.35}\\
\midrule
\multirow{2}{*}{`60 - 80'}&TalkNet&96.95&93.16&14.37\\
&\textbf{TS-TalkNet}&\textbf{97.30}&\textbf{93.99}&\textbf{12.97}\\
\midrule
\multirow{2}{*}{`80 - 100'}&TalkNet&99.71&93.44&13.89\\
&\textbf{TS-TalkNet}&\textbf{99.75}&\textbf{94.01}&\textbf{13.28}\\
\bottomrule
     \end{tabular}}
 \label{tab:analy}
 \vspace{-0.5cm}
\end{table} 

\vspace{-1mm}
\subsection{Qualitative studies}
\label{subsection:study}
\textbf{Fusion structure.}
We implement three fusion approaches to combine the audio-visual embeddings and speaker embedding, as mentioned in Subsection~\ref{subsection:fusion}. As shown in Table~\ref{tab:main2}, for TS-TalkNet~(Fus1), we obtain the improvements by 0.9\%, 0.4\% and 0.6\% in terms of AP, AUC and EER, respectively, over the TalkNet on the ASW val set; and 0.6\%, 0.3\% and 0.3\% in terms of AP, AUC and EER, respectively, on the ASW test set. For TS-TalkNet~(Fus2), we achieve the improvements by 1.1\%, 0.5\% and 0.9\% in terms of AP, AUC and EER, respectively, over TalkNet on the ASW val set; and 0.6\%, 0.3\% and 0.7\% in terms of AP, AUC and EER respectively on the ASW test set. Similar observations can be found in Table~\ref{tab:main1}. 

These results indicate that incorporating the speaker embedding can improve the performance of TS-TalkNet, regardless of the fusion approaches. Therefore, our idea of using speaker characteristics to assist ASD is not restricted to a specific network architecture.
~\\
\textbf{Analysis study.}
We aim to investigate the effect of speaker embedding incorporation for face tracks with varying proportions of active frames. So we analyze the performance improvement of TS-TalkNet on the ASW val set by dividing the instances into five sections based on the percentage of the frames with active labels of each track input.
As shown in Table~\ref{tab:analy}, TS-TalkNet can improve the detection accuracy for each section, indicating that the speaker-specific characteristics are valuable for detecting the target speaker's active frames across different scenarios.

\vspace{-0.5cm}
\section{Conclusion}
In this paper, we propose the Target Speaker TalkNet (TS-TalkNet) for active speaker detection by incorporating target speaker embeddings. Our experiments and studies demonstrate that speaker characteristic plays a crucial role in solving the ASD problem. Our work provides a new perspective and potential solution for improving the performance of ASD systems in complicated acoustic environments. In future work, we can investigate the potential applications of TS-TalkNet in other speech-related tasks.

\section{Acknowledgements}
This work is funded by
1) Huawei Noah’s Ark Lab;
2) National Natural Science Foundation of China (Grant No. 62271432);
3) Agency for Science, Technology and Research (A*STAR) under its AME Programmatic Funding Scheme (Project No. A18A2b0046)

\newpage
% \pagebreak
\balance
\bibliographystyle{IEEEtran}
\bibliography{main}

\end{document}